\documentclass[final]{aipproc}
\usepackage{graphics,graphicx,amsmath,array,hhline,fancyhdr,mathptm}
\usepackage[english]{babel}
\def\be{\begin{equation}}
\def\ee{\end{equation}}
\def\bea{\begin{eqnarray}}
\def\eea{\end{eqnarray}}

\def\d{{\mbox{\rm d}}}

\layoutstyle{6x9}

\begin{document}

\title{Bose-Einstein or HBT correlation signature of a second order QCD phase transition}

\classification{}
\keywords      {}

\author{T. Cs\"org\H{o}$^*$, S. Hegyi}
{address={MTA KFKI RMKI, H - 1525 Budapest 114, P.O.Box 49,
Hungary}}

\author{T. Nov\'ak}
{address={University of Nijmegen, NL - 6525 ED Nijmegen, Toernooiveld 1
} }

\author{W. A. Zajc}
{address={Department of Physics, Columbia University, 538 W 120th Street, New York, NY 10027, USA
}}

\begin{abstract}
For particles emerging from a second order QCD phase transition, 
we show that a recently introduced shape parameter of the Bose-Einstein correlation function, 
the L\'{e}vy index of stability equals to the correlation exponent - 
one of the critical exponents that characterize the behavior of the matter 
in the vicinity of the second order phase transition point.
Hence the shape of the Bose-Einstein / HBT correlation functions, 
when measured as a function of bombarding 
energy and centrality in various heavy ion reactions, 
can be utilized to locate experimentally the second order phase transition and 
the critical end point of the first order phase transition line in QCD.
\end{abstract}

\maketitle
\section{Introduction}
        The study fractal phenomena was initiated in high energy particle and nuclear physics by
        Bialas and Peschanski in ref.~\cite{Bialas:1985jb}, with the motivation of searching 
	for a second order phase transition by studying intermittency or
	the the power-law behavior of moments of the multiplicity
	distribution in narrowing bins of the momentum space, see refs.
	~\cite{Bialas:1990gt,DeWolf:1995pc} for excellent reviews on this topic.

        The mathematical properties of Bose-Einstein correlation functions
        for L\'{e}vy stable sources were written up by three of us in
        refs.~\cite{Csorgo:2003uv,Csorgo:2004ch}, and are
	recapitulated in the next section.
        In ref.~\cite{Csorgo:2004sr}
	 we have added a physical interpretation and showed,
        that in case of jet physics, 	
	the fractal properties of QCD cascades can naturally  be measured
        by the L\'{e}vy index of stability of the Bose-Einstein correlation functions.
        Our analytic results were similar in spirit to the numerical investigations
        of Wilk and collaborators in ref.~\cite{Utyuzh:1999zg}. Note that these correlations 
	are frequently referred to as Hanbury Brown - Twiss or HBT correlations in	
	the literature of heavy ion physics.
	
        Bialas realized, that  Bose-Einstein correlations and
        intermittency might be deeply connected~\cite{Bialas:1992ca},
	and considered a distribution of Gaussians where the radius
	parameter of the Gaussian has a power-law distribution, thus
	giving a way to the study fractals in coordinate space with the
	help of Bose-Einstein correlations. 
	Brax and Peschanski were the first to introduce L\'{e}vy
	distributions, in momentum space, to multiparticle production in high 
	energy physics~\cite{Brax:1990jv}. They have
	suggested to use the measured value of the L\'evy index of stability to
	signal quark gluon plasma production in heavy ion physics.
	Here we reconnect these seemingly different topics, and show how
	the excitation function of the shape parameter of the correlation
	function can be utilized to locate experimentally the critical end-point of QCD.

\section{Bose-Einstein correlations \& L\'evy stable sources}
	The two-particle Bose-Einstein correlation function
        is defined with the help of the two-particle and single-particle invariant
        momentum distributions as:
\begin{equation}
	C_2({\mathbf k}_1,{\mathbf k}_2) = 
		\frac{N_2({\mathbf k}_1,{\mathbf k}_2)}
		{N_1({\mathbf k}_1)\, N_1({\mathbf k}_2)}.  \label{e:cdef}
\end{equation}
        If long-range correlations can be neglected or corrected for,
        and if the short-range correlations are dominated by Bose-Einstein
        correlations, this two-particle Bose-Einstein correlation function
        is related to the Fourier-transformed source distribution.
        For clarity, let us consider the case of a one-dimensional,
        factorized source, 
\begin{equation}
        S(x,k) =  f(x) \, g(k).
\end{equation}
	 In this case~\cite{Csorgo:2003uv,Csorgo:2004ch},
        the Bose-Einstein correlation function is
\begin{equation}
        C_2(k_1,k_2) = 1 + |\tilde f(q)|^2,
\end{equation}
        where the Fourier transformed source density (often referred to as the
        {\it characteristic function}) and the relative momentum are defined as
\begin{equation}
        \tilde f(q)  =
                \int  \mbox{\rm d}x \, \exp(i q x) \,f(x),\qquad\quad 
        q  =  k_1 - k_2 .\label{e:fourier}
\end{equation}

	For the case of the jets decaying to jets to jets and so on, 
	as well as at a second order phase transition, where fluctuations appear
	on all possible scales with a power-law tailed distribution, the final position
        of a particle is given by a large number of position shifts,
        hence the distribution of the final position $x$ is obtained as  a convolution,
\be
        x = \sum_{i=1}^n x_i, \qquad\qquad
                f(x) = \int \Pi_{i=1}^n \d x_i\, \Pi_{j=1}^n f_j(x_j)\,
            \delta(x - \sum_{k=1}^n x_k ).
\ee
        In the case of particle emission from QCD jets, that the
        fractal defining the particle emission is infrared stable:
        adding one more, very soft gluon does not change the
        resulting source distributions. 
	A similar property holds for systems at a second order phase transition:
	the system becomes invariant under a renormalization group transformation.
	Bose-Einstein correlation functions for such particle
        emitting sources were evaluated recently by three of us,
        which we summarize here
	following refs.~\cite{Csorgo:2003uv} and ~\cite{Csorgo:2004ch}.

	Various forms of the Central Limit Theorem state,
        that under certain conditions, the distribution of the sum of
        large number of random variables converges (for $n \rightarrow \infty$)
        to a limit distribution.
        In case of ``normal" elementary processes, that have finite means
        and variances, the limit distribution of their sum is a Gaussian.
	In case of random motion in a thermal medium, such position
	distribution corresponds to normal diffusion. However, near a
	second order phase transition point, fluctuations appear on all
	scales and the variance of the elementary process diverges, 
	corresponding to the so-called anomalous diffusion.
	In this case, the system still may be invariant under convolution, 
	and the shape of the limit distribution
	becomes independent from the number of elementary steps.

\begin{figure}
\includegraphics[angle=0,width=0.5\textwidth]{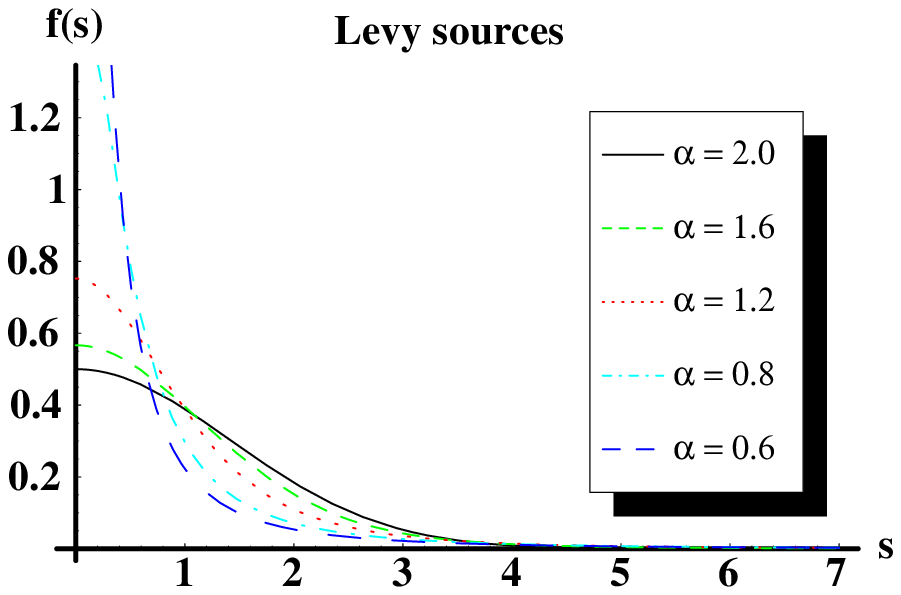}
\includegraphics[angle=0,width=0.5\textwidth]{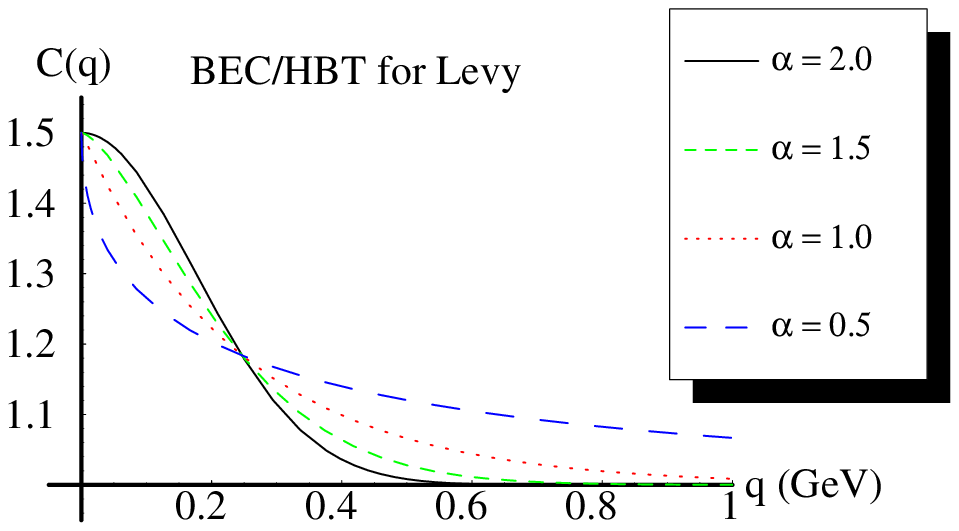}
\caption{ 
{\small (left)
Source functions for univariate symmetric L\'{e}vy laws, as
a function of the dimensionless variable $s = r/R $, 
for various values of the L\'{e}vy index of stability, $\alpha$.
(right) Bose-Einstein correlation (or HBT) correlation functions
for univariate symmetric L\'{e}vy laws, for a fixed  scale parameter of $R = 0.8$ fm
and various values of the L\'{e}vy index of stability, $\alpha$.
}
}
\label{f:fig1}
\end{figure}

        Stable distributions are precisely those limit distributions
        that can occur in Generalized Central Limit theorems. Their study
        was begun by the mathematician Paul L\'{e}vy in the 1920's.
        The stable distributions can be given in terms of
        their characteristic functions, as the Fourier transform of
        a convolution is a product of the Fourier-transforms,
    \be
            \tilde f(q) = \prod_{i=1}^n \tilde f_i(q) ,
    \ee
        and limit distributions appear when the convolution
        of one more elementary process does not change the
        shape of the limit distribution,  but it results only in
        a modification of the parameters of the limit distribution.
        The characteristic function of univariate and symmetric stable distributions is
\be
        \tilde f(q)=\exp\left( i q \delta -|\gamma q|^\alpha\right), \label{e:fqs}
\ee
        where the support of the density function $f(x)$ is $(-\infty,\infty)$.
        Deep mathematical results imply that the index of stability, $\alpha$,
        satisfies the inequality $0 < \alpha \le 2$,
        so that the source distribution be always positive.
        These L\'{e}vy distributions are indeed stable under convolutions,
        in the sense of the following relations:
\bea
        \tilde f_i(q) & = & \exp\left( i q \delta_i -|\gamma_i q|^\alpha\right), \qquad
        \prod_{i=1}^n \tilde f_i(q) \, =  \,
        \exp\left( i q \delta -|\gamma q|^\alpha\right) ,\\
        \gamma^\alpha & = &  \sum_{i=1}^n \gamma_i^\alpha,
        \quad\qquad\qquad\qquad
        \delta \, = \, \sum_{i=1}^n \delta_i. \label{e:gami}
\eea

Thus the Bose-Einstein correlation functions for uni-variate,
        symmetric stable distributions (after a core-halo correction, and a re-scaling) read as
\begin{equation}
        C(q;\alpha) = 1 + \lambda \exp\left(-|q R|^\alpha\right).  \label{e:BEC-Levy}
\end{equation}
        Refs.~\cite{Csorgo:2003uv} and ~\cite{Csorgo:2004ch} discuss further examples and details
        and generalize these results to
        three dimensional, hydrodynamically expanding, core-halo type sources, as well
	as to three-particle correlations. 

\section{The anomalous dimension of  QCD jets and BE/HBT }
                                                                                                                                                                                                                                                                                               
	In QCD, jets emit jets that emit additional jets and so on.
        The resulting fractal structure of QCD jets was related to
	intermittency and power-law dependence of multiplicity moments
	on the bin-size in momentum space with the help of a beautiful 
	geometric interpretation of the color dipole picture 
	in refs.~\cite{Dahlqvist:1989yc,Gustafson:1990qi,Gustafson:1990qk},
	and an infrared stable measure on the parton states,
	related to the hadronic multiplicity distribution.
        These ideas were developed further in refs.
        ~\cite{Gustafson:1991ru,Gustafson:1992uh,Andersson:1995jv},  
	~\cite{Dokshitzer:1992df} as well as in refs.~\cite{Ochs:1992tg,Ochs:1994rt}.

\begin{figure}[!thb]
\includegraphics[angle=0,width=0.8\textwidth]{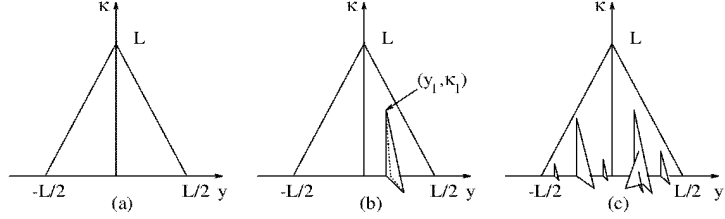}
\label{fig:fractal-jets}
\caption{
{\small The phase-space of QCD jets in the $(y,\kappa)$ 
 	plane, where $\kappa = \log(k_t^2)$.
	(a) The phase space available for a gluon emitted by a high energy $q\overline{q}$
	system is a triangular region in the $(y,\kappa)$ plane. 
	(b) If one gluon is emitted at $(y_1,\kappa_1)$,
   	the phase space for a second (softer) gluon is
	given by the area of this folded surface. 
	(c) The total gluonic phase space can be described by this
	multifaceted fractal surface~\cite{Dahlqvist:1989yc,Gustafson:1990qi,Gustafson:1990qk}.
} 
}
\label{f:fig2}
\end{figure}
	
	A high energy $q\overline{q}$ system radiates gluons according to the dipole formula    
\begin{equation}
  dn=\frac{3 \alpha_s}{4\pi^2} \frac{ d k_\perp^2}{ k_\perp^2} dy d\phi, 
	\label{e:dn}
\end{equation}
	hence the phase-space for the emission of a gluon is given by the relation
\begin{equation}
	| y| \le \frac{1}{2} \ln(s/k_\perp^2),
\end{equation}
	 which corresponds to the triangular region in a $(y,\ln k_\perp^2)$ diagram
 	as shown in Fig. ~\ref{f:fig2} (a). 
	If two gluons are emitted, then the distribution of the hardest gluon
 	is described by eq. (\ref{e:dn}). 
	The distribution of the second, softer, gluon corresponds to
 	two dipoles, the first is stretched between the quark and the first gluon, and 
 	the second between the first gluon and the anti-quark. 
 	The phase-space available for the second gluon corresponds to the
 	folded surface in Fig. ~\ref{f:fig2} (b), with the constraint $k_{\perp,2}^2 < k_{\perp,1}^2$,
 	as the first gluon is assumed to be the hardest one.
	This procedure can be generalized so that the emission of a third, still softer
	gluon corresponds to radiation from three color dipoles, with $n$ gluons emitted already
       	the emission of the $n+1$-th gluon is given by a chain of $n+1$ dipoles. 
	Thus, with many gluons, the gluonic phase space can be represented by a 
	multi-faceted surface as illustrated in Fig. ~\ref{f:fig2} (c). 
	Each gluon adds a fold to the surface, which increases the phase-space for softer gluons.
	(Note, that in this process the recoils are neglected, 
	as is normal in leading log approximation).
	Due to its iterative nature, the process generates 
	a Koch-type fractal curve at the base-line.
	The length of this base-line of the partonic structure on Figure~\ref{f:fig2} (c) 
	is proportional to the particle multiplicity. 
	This curve is longer, when studied with higher resolution: 
	it is a fractal curve, embedded into the four-dimensional
	energy-momentum space, characterized by the fractal dimension
\be 
    d_f = 1 + \sqrt{\frac{3 \alpha_s}{2 \pi}}, 
\ee
	or one plus the anomalous dimension 
	of QCD~\cite{Dahlqvist:1989yc,Gustafson:1990qi,Gustafson:1990qk}.
    With the help of the Lund string fragmentation picture, this fractal in momentum space is mapped
    into a fractal in coordinate space,  and the constant of conversion is
    the hadronic string tension, $\kappa \approx 1$ GeV/fm. This
    mapping does not change the fractal properties of the curve.

	A walk, where  the length of the steps is given by a L\'evy distribution,
        and the direction of the steps is random, corresponds to a fractal curve,
	in physical terms it can be interpreted as the path of a test particle
	performing a generalized Brownian motion. This motion is referred to
	as anomalous diffusion and the probability that the test particle diffuses
	to distances $r$ greater than a certain value of $|s| $ is given by
	$P(r > |s|)  \propto |s|^{-\alpha}$. This relation is valid for anomalous
	diffusion in any dimensions. 
	Thus the L\'evy index of stability $\alpha$ is the fractal dimension
	of the trajectory of the corresponding anomalous diffusion~\cite{Seshadri82ab}.
	When we apply this result to QCD, there are two key considerations.
       
       	First, if gluon radiation is neglected, the $q\overline{q}$ system
	hadronizes as a 1+1 dimensional hadronic string, which has no fractal
	structure. If the gluon emission is switched on, the emission of gluon $n$ from 
       one of the $n$ dipoles 	corresponds to a step of an anomalous diffusion 
	in the plane transverse to the given  dipole. Hence the anomalous 
	dimension of QCD equals to the L\'evy index of stability 
	of this anomalous diffusion,	
\be
	\sqrt{\frac{3 \alpha_s}{2\pi}} = \alpha_{\mbox{\tiny\rm L\'evy}}  . 
\ee	
	Second, data on Bose-Einstein correlations are often determined
       in terms of the invariant momentum difference $Q_{\rm inv} = \sqrt{-(p_1-p_2)^2} $.
	Bose-Einstein correlation functions that depend on this invariant momentum
       difference can be obtained within the framework of the so-called 
	$\tau$-model.       This model assumes a broad proper-time distribution,
	$H(\tau)$ and very strong correlations
	between coordinate and momentum in all directions, $x^\mu/\tau \propto p^\mu/m_{(t)}$.
	Hence $(x_1 - x_2) (p_1 - p_2) \propto \tau Q_{\rm inv}^2$,
	see refs.~\cite{cstjz,csorgo-alushta} for details.
	In this case, the Bose-Einstein correlation function measures the
	Fourier-transformed proper-time distribution $\tilde H$ in the following,
	unusual manner:
\be
	C_2(Q_{\rm inv}) \simeq 
	1 + \lambda \, \mbox{\it Re}\, \tilde H^2\left(\frac{Q_{\rm inv}^2}{2  m_{(t)}}\right),
\ee
	where $m_{(t)} $ stands for the (transverse) mass of the pair
	for (two)- or more jet events.	
	From this relation it follows, that $\alpha_{\mbox{\tiny\rm BEC}} = 
	2 \alpha_{\mbox{\tiny\rm L\'evy}}$.
	Thus we find the following relationship between the
	running QCD coupling constant $\alpha_s$ and the exponent of an invariant relative 
	momentum dependent Bose-Einstein correlation function $\alpha_{\mbox{\tiny\rm BEC}}$:
\be
	\alpha_s = \frac{\pi}{6} \alpha^2_{\mbox{\tiny\rm BEC}}.\label{e:QCDa}
\ee

	In ref.~\cite{Csorgo:2004sr} we have compared this leading log result to 
	 NA22 and UA1 correlation data of refs.
	~\cite{Agababian:sn},\cite{Neumeister:1993bf}
	and found a reasonable agreement with these data.

\section{Bose-Einstein correlations at a  second order QCD phase transition}
	The main motivation behind the experimental and theoretical program
	of high energy physics is to study the phase diagram of hot and dense
	hadronic matter. According to recent lattice QCD calculations at finite
	temperature and baryon density, there exist a line of first order
	phase transitions that separates the hadronic and the quark-gluon plasma (QGP) state.
	This line of the first order phase transitions ends at the critical
	end-point (CEP), where the transition from hadron gas to QGP becomes
	a second order phase transition. Recent lattice QCD calculations
	located~\cite{Fodor:2004nz}
	this CEP at $T_E = 162 \pm 2$ MeV and
	$\mu_E = 360 \pm 40$ MeV. Below these baryochemical values,
	the transition from a hadron gas to a QGP becomes a 
	cross-over, and at vanishing net baryon density
	the critical temperature becomes $T_c = 164 \pm 2 $ MeV  (the errors
	are statistical only). In this calculation, the quark masses were already at the
	physical value, but the continuum extrapolation was missing.
	S. Katz presented improvements at the Quark Matter 2005
	conference~\cite{Katz:2005br}, using physical quark masses and working towards the 
	continuum extrapolation. He reported $T_c = 189 \pm 8$ MeV 
	for the critical temperature at $\mu_B = 0$. 

	At the CEP, the second order phase transition is characterized by
	the fixed point of the renormalization group transformations.
	In a quark-gluon plasma, the vacuum expectation value of the
	quark condensate $ c = \langle \overline q \, q\rangle$ vanishes ,
	while in the hadronic phase, this vacuum expectation value becomes
	non-zero.  The correlation function of the order parameter is defined as
	$\rho(R) = \langle c(r+R) c(r) \rangle - \langle c\rangle^2 $
	and measures the spatial correlation between the pions.
	At the CEP, this correlation function decays as
\be
	\rho(R) \propto R^{- (d - 2 + \eta)} ,
\ee
	 a power-law. The parameter $\eta$ is called as the exponent of the correlation function.

	For L\'{e}vy stable sources, corresponding to an anomalous diffusion
	with large fluctuations in coordinate space, the correlation between the
	initial and actual positions decays also as a power-law, where the
	exponent is given by the L\'{e}vy index of stability $\alpha$ as
\be
	\rho(R) \propto R^{- (1 + \alpha)} .
\ee
	As we are considering a QCD phase transition in a $d=3$ three-dimensional coordinate space
	we find that the correlation exponent equals to the L\'{e}vy index of stability,
$
	\alpha = \eta .
$
	Stephanov, Rajagopal and Shuryak pointed out~\cite{Stephanov:1998dy},
	that the universality class of the 
	second order QCD phase transition is that of the 3d Ising model.
	For this universality class, the correlation exponent has been determined
	by Rieger ~\cite{Rieger:1995ah} as
\be
	\alpha(\mbox{\rm L\'evy}) = \eta(\mbox{\rm 3d Ising}) = 0.50 \pm 0.05.
\ee
	Fig. 1 indicates that the change in the shape of the correlation function
	is rather significant, if $\alpha$ decreases from its Gaussian value of 2
	to 0.5, its characteristic value at the 2nd order QCD phase transition point.
	Fig. 3 illustrates how this shape parameter of the Bose-Einstein correlation
	function may depend on the relative temperature near the critical point.
	Hence studying the {\it shape} parameter of the two-particle
	Bose-Einstein or HBT correlation functions as a function of the
	bombarding energy or the centrality of the heavy ion collisions,
	a previously unknown tool is obtained to determine if the pions
	are emitted from the neighborhood of the critical end point of the
	QCD phase diagram. 

	Furthermore, based on an universality class argument, we have
	determined that the second order  QCD phase transition at the critical
	end point will be signaled with
	the  value of     $\alpha = 0.5$, a very spiky Bose-Einstein correlation
	function indeed.  
 
\begin{figure}[ht]
\includegraphics[angle=0,width=0.8\textwidth]{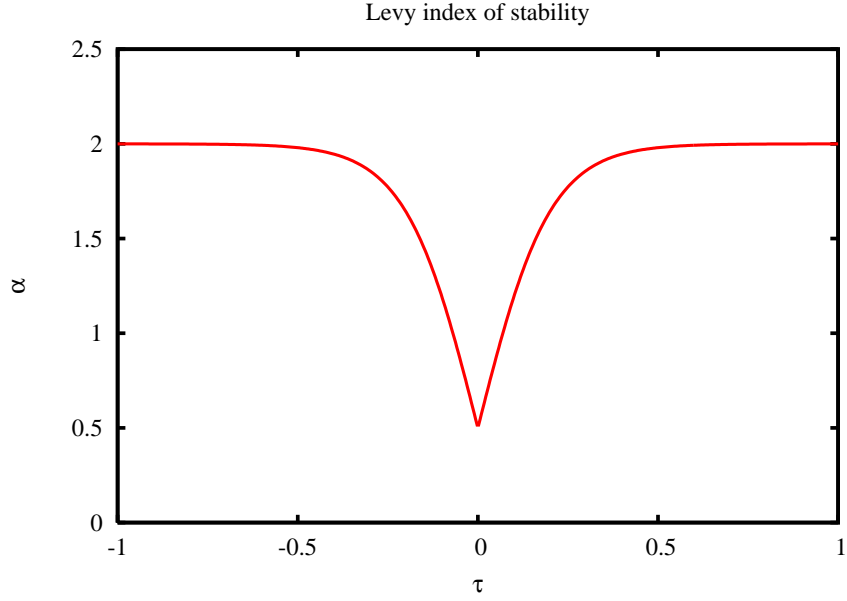} \\
\caption{
{\small Illustration of the behavior of  the L\'evy index of stability 
	of Bose-Einstein correlations as a function of the dimensionless temperature
	variable $\tau = (T-T_c)/T_c$ in the neighborhood of the critical endpoint
	of the 1st order phase transition line in QCD. At the critical endpoint, the
	phase transition becomes 2nd order and the L\'evy index of stability
	decreases to the correlation exponent of QCD. As this transition has the
	same universality class as that of the 3d Ising model, one expects
	a decrease from the $\alpha \approx 2$ values that are characteristic
	to a Boltzmann gas and normal diffusion to $\alpha = 0.5 $,
	corresponding to the correlation exponent of QCD at the critical endpoint.
	As shown in Fig. 1, such a change in the shape parameter makes the
	Bose-Einstein correlation functions much sharper than a simple
	Gaussian, so the spiky structure of the correlation function could be	
	used to search for this point experimentally.
}
	}
  \label{f:fig3}
\end{figure}

\section{Conclusions}

	We have recapitulated earlier results that indicate, that the general shape
	of the Bose-Einstein or HBT correlation functions is a stretched exponential
	or L\'evy stable form, where the L\'{e}vy index of stability becomes
	 a new shape parameter of the correlation function with $0 < \alpha \le 2$
	and the popular Gaussian parameterization corresponds to the  $\alpha = 2$
	particular, special case. Then  we have studied two physically interesting examples. 

	In case of particle emission from jets, we have recapitulated 
	the connection between the stability index of the Bose-Einstein/HBT correlation functions 
	and the running coupling constant of QCD.

	We have also considered a scenario, when the power-law tail of a L\'{e}vy distribution
	of the particle emission {\it in the coordinate space} appears due to
	a second-order QCD phase transition.  
	In this case, the L\'{e}vy index of stability of the Bose-Einstein or HBT 
	correlation function was shown to be equal to the correlation exponent of QCD. 
	This value is known to be $0.5 \pm 0.05$ from universality class considerations.
	 Hence by measuring the excitation function of 
	the L\'{e}vy index of stability (the shape parameter
	of the two-particle Bose-Einstein or  HBT correlation functions), 
	one can experimentally determine the bombarding energy and centrality range where
	a heavy ion collision hits the critical end point of QCD.
	Clearly, more work is necessary to check to what extent this interesting effect 
	can be masked by the decays of various resonances, hydrodynamic expansion,
	and by the time evolution of the particle emitting source between 
	the second order phase transition point  and the freeze-out temperature.

\begin{theacknowledgments}
It is our pleasure to acknowledge the inspiring discussions
with A. Bialas, R. Glauber, W. Kittel  and R. Peschanski.
This research was supported by the NATO Collaborative Linkage Grant PST.CLG.980086,
by the Hungarian - US MTA OTKA NSF grant INT0089462
and by the OTKA grants T038406 and T049466.
\end{theacknowledgments}

\bibliographystyle{aipprocl} 


\end{document}